\documentclass[preprint,12pt]{elsarticle}

\usepackage{amssymb}
\usepackage{multirow}
\usepackage{booktabs}
\usepackage{amssymb}
\usepackage{mathrsfs}
\usepackage{longtable}
\usepackage{lscape}
\usepackage{tabularx}
\usepackage{epsfig}
\usepackage{colortbl}
\usepackage{subfigure}
\journal{Physica A}

\begin{document}

\begin{frontmatter}



\title{Tsallis information dimension of complex networks}


\author[swu]{Qi Zhang}
\author[swu]{Meizhu Li}
\author[swu,NWPU,vu]{Yong Deng\corref{cor}}
\ead{ydeng@swu.edu.cn, prof.deng@hotmail.com}
\author[vu]{Sankaran Mahadevan}

\cortext[cor]{Corresponding author: Yong Deng, School of Computer and Information Science, Southwest University, Chongqing, 400715, China.}

\address[swu]{School of Computer and Information Science, Southwest University, Chongqing, 400715, China}
\address[NWPU]{School of Automation, Northwestern Polytechnical University, Xian, Shaanxi 710072, China}
\address[vu]{School of Engineering, Vanderbilt University, Nashville, TN, 37235, USA}
%\address[UB]{School of Engineering, University of British Columbia Okanagan, 3333 University Way, Kelowna, BC, Cannada}

\begin{abstract}
The fractal and self-similarity properties are revealed in many complex networks. In order to show the influence of different part in the complex networks to the information dimension, we have proposed a new information dimension based on Tsallis entropy namely Tsallis information dimension. The Tsallis information dimension can show the fractal property from different perspective by set different value of $q$.

%The fractal and self-similarity properties are revealed in many complex networks, the dimension represents the fractal and self-similarity properties of complex networks. The information dimension use the information of the structure to describe the fractal and self-similarity properties of the complex networks. In the information dimension of complex networks, the nodes have more connection with others is important. In fact, no matter how many nodes contains in the box, every box can influence in the information dimension. The classical method defines the effect of every box based on the probability of nodes distribute. The main effect is constituted by the probabilities which have large value. However, sometimes the small value probabilities also play an important role in the information dimension. To reveal the relationship between information dimension and those different kind of probabilities. Based on the Tsallis entropy and the box-covering algorithm, we propose a new information dimension of complex networks, named Tsallis information dimension. It is applied to calculate the fractal dimensions of complex networks in different perspective by focus on different probabilities, it have revealed the fractal property of the complex networks in comprehensive. The results show that the fractal property of the complex networks is a dynamic quantity.
\end{abstract}
\begin{keyword}
Complex networks \sep Information dimension \sep Tsallis entropy \sep Tsallis information dimension

%% or \MSC[2008] code \
% code (2000 is the default)

\end{keyword}
\end{frontmatter}

\section{Introduction}
\label{Introduction}
The complex networks have been applied in many disciplines \cite{newman2003structure,yu2009pinning,song2010synchronization,hu2011synchronization,vidal2011interactome,wei2013networks,zhang2013impulsive}.  Researchers have revealed several properties of the complex networks, such as small-world phenomena \cite{watts1998collective}, scale-free degree \cite{barabasi1999emergence}, fractal, self-similarity and community structure \cite{fortunato2010community},etc. The fractal and self-similarity properties have shown the structure characteristic of the complex networks, many researchers have been attracted to explore it \cite{mandelbrot1984fractal,locci2010three,moret2012classical,smolyaninov2012metamaterial,redelico2012empirical,rajagopalan2013brain,chelminiak2013emergence}. In order to describe the fractal properties, Song ${et.al}$ proposed the dimension of the complex networks \cite{song2005self,song2006origins,song2007calculate}.

Recently, an information dimension of the complex networks has been proposed by Wei ${et.al}$ in \cite{wei2014informationdimension}. In the information dimension, the boxes which contain more nodes have a maximum effect to the information dimension. However, sometimes those boxes contain few nodes may play an important role in the fractal property. In order to show the influence of the boxes which have different mounts of nodes to the information dimension. A new information dimension based on Tsallis entropy \cite{tsallis1988possible} is proposed in this paper. In the proposed method, setting different values of ${q}$ means chose different part as the main effect of the information dimension.

The rest of this paper is organised as follows. Section \ref{Rreparatorywork} introduces some preliminaries of this work. In section \ref{Tsinformationd}, a new information dimension of complex networks based on the Tsallis entropy is proposed. The application of the proposed method is illustrated in section \ref{application}. Conclusion is given in Section \ref{conclusion}.
.
\section{Preliminaries}
\label{Rreparatorywork}
\subsection{Box-covering algorithm of complex networks}
\label{Boxcovering}
Song ${et.al}$ have proposed a new box-covering algorithm for complex networks \cite{song2007calculate,song2006origins,song2005self}. It contains a new definition for the box size ${l_B}$ which is based on the distances between the nodes in the complex networks.
\begin{figure}[!ht]
\begin{center}
\includegraphics[width=9cm,height=6cm]{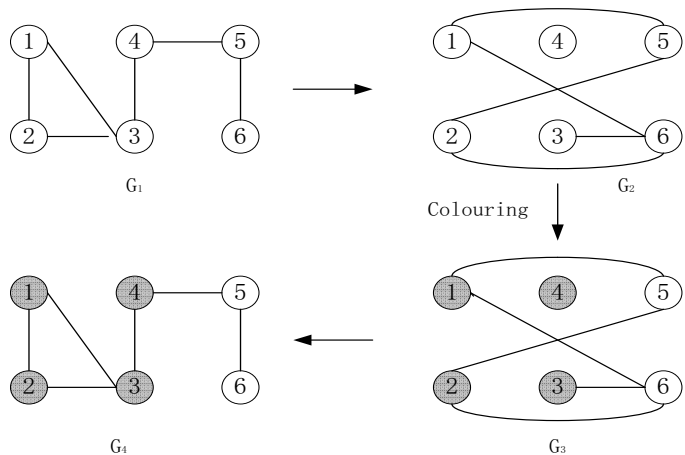}
\caption{The classical box-covering algorithm for complex networks, where $l=3$. The network $G_1$ is original network with 6 nodes and 6 edges. The network $G_2$ is obtained by only connecting to nodes which distance between them not less than 3 in network $G_3$. The network $G_3$ is obtained when the greedy algorithm is used for node coloring on $G_2$ \cite{song2007calculate}.}
\label{fp1}       % Give a unique label
\end{center}
\end{figure}

For a given network $G_1$ and box size ${l_B}$, a box is a set of nodes where all distances ${l_{ij}}$ between any two nodes $i$ and $j$ in the box are smaller than ${l_B}$, The minimum number of boxes required to cover the entire networks is denoted by ${N_B}$.

For ${l_B}$=1, ${N_B}$ is obviously equals to the size of the network ${N}$, while ${N_B}$=1 for ${l_B\geq{l_B^{\max }}}$, where ${l_B^{\max }}$ is the diameter of the network plus one, the diameter of the network equals to the maximum distance in the network \cite{song2007calculate}.

If the distance between two nodes in ${G_1}$ is greater than ${l_B}$, these two neighbors cannot belong in the same box. According to the construction of ${G_2}$, these two nodes will be connected in ${G_2}$ and thus they will not belong in the same box in ${G_1}$. On the contrary, if the distance between two nodes in ${G_1}$ is less than ${l_B}$, it is possible that these nodes belong in the same box. In ${G_2}$ these two nodes will not be connected and it is allowed for these nodes to carry the same color, it will belong to the same box in ${G_1}$ \cite{song2007calculate}. More details are shown in Fig. \ref{fp1}.

 The box-covering algorithm is used to calculate the minimum numbers of box ${N_l}$ by Song ${et.al}$. Then the fractal dimension ${d_B}$ of the complex networks can be described by the relationship between ${N_l}$ and ${l_B}$. The details are shown in Eq. (\ref{db1}).
\begin{equation}
\label{db1}
{d_B}{\rm{ =  - }}\mathop {\lim }\limits_{{l_B} \to 0} \frac{{\ln {N_l}}}{{\ln {l_B}}}
\end{equation}

\subsection{Tsallis entropy}
\label{TSentropy}
The entropy is defined by Clausius for thermodynamics \cite{clausius1867mechanical}, connects the macroscopic and microscopic worlds. For a finite discrete set of probabilities the definition of the Boltzmann-Gibbs \cite{gibbs1902elementary} entropy is given as follows:

\begin{equation}\label{S_BG}
{S_{BG}} =  -k \sum\limits_{i = 1}^N {{p_i}} \ln {p_i}
\end{equation}

Where ${BG}$ stands for ${Boltzmann-Gibbs}$, the ${S_{BG}}$ represents the Boltzmann-Gibbs entropy. The conventional constant $k$ is the Boltzmann universal constant for thermostatistical systems, the value of $k$ will being taken to be unity in information theory \cite{tsallis2010nonadditiveJPCS,shannon2001mathematical}.

In 1988, a more general form for entropy have been proposed by Tsallis  \cite{tsallis1988possible}. It is shown as follows:

\begin{equation}\label{S_q}
{S_q} =  -k \sum\limits_{i = 1}^N {{p_i}} {\ln _q}{p_i}
\end{equation}

The $q-logarithmic$ function in the Eq. (\ref{S_q}) is presented as follows \cite{tsallis2010nonadditiveJPCS}:
\begin{equation}\label{ln_q}
{\ln _q}{p_i} = \frac{{{p_i}^{1 - q} - 1}}{{1 - q}}({p_i} > 0;q \in \Re ;l{n_1}{p_i} = ln{p_i})
\end{equation}

Based on the Eq. (\ref{ln_q}), the Eq. (\ref{S_q}) can be rewritten as follows:
\begin{equation}\label{S_q1}
{S_q} = k\frac{{1 - \sum\limits_{i = 1}^N {{p_i}^q} }}{{q - 1}}
\end{equation}

Where ${N}$ is the number of the subsystems.

%The additivity between two independent systems ${f(x)}$ and ${f(y)}$ can be described by the value of $q$ \cite{tsallis2009nonadditiveEPJA}. The details is revealed in Table \ref{tab-q}.
%\begin{table}[!ht]
%\begin{center}
%\caption{The influence of ${q}$ in the system}
%\label{tab-q}
%\begin{tabular}{ll}
%\hline
%Scope of q value&nonextensive property \\
%\hline
%${q<1}$ & ${f(x + y) \ge f(x) + f(y)}$ \\
%${q=1}$ & ${f(x + y){\rm{ = }}f(x) + f(y)}$ \\
%${q>1}$ & ${f(x + y) \le f(x) + f(y)}$ \\
%\hline
%\end{tabular}
%\end{center}
%\end{table}

%The Tsallis entropy have described the different additivity among the subsystems, it can be used to describe the relationship between fractal property and information dimension \cite{tsallis2009introduction}.
\subsection{Information dimension}
\label{informationdimension}
Based on the information entropy and the box-covering algorithm, an information dimension has been proposed by Wei ${et.al}$ in \cite{wei2014informationdimension}.

The information of the complex networks is shown as follows:
\begin{equation}
\label{Infor}
  I =  - \sum\limits_{i = 1}^{{N_b}} {{p_i}\ln ({p_i})}
\end{equation}

The ${p_i}$ in the Eq. (\ref{Infor}) represents the probability of the nodes in the $i$th box. It is shown in Eq. (\ref{pi}).

\begin{equation}
\label{pi}
{p_i} = \frac{{{n_i}}}{n}
\end{equation}

Where ${n_i}$ is the node number in the $i$th box, $n$ is the total number of the nodes in the complex networks \cite{wei2014informationdimension}.

Depends on the relationship between information of the complex networks and the box size. The information dimension of the complex networks is shown in Eq. (\ref{indimension}) \cite{wei2014informationdimension}.
\begin{equation}
\label{indimension}
{d_b} =  - \mathop {\lim }\limits_{l \to 0} \frac{{I}}{{\ln l}} = \mathop {\lim }\limits_{l \to 0} \frac{{\sum\limits_{i = 1}^{{N_b}} {{p_i}\ln ({p_i})} }}{{\ln l}}
\end{equation}

Where ${d_b}$ is the information dimension of the complex network. Based on Eq. (\ref{Infor}), the Eq. (\ref{indimension}) can be rewritten as follows:
\begin{equation}
\label{informationdimension}
{d_b} = \mathop {\lim }\limits_{l \to 0} \frac{{\sum\limits_{i = 1}^{{N_b}} {\frac{{{n_i}(l)}}{n}\ln (\frac{{{n_i}(l)}}{n})} }}{{\ln l}}
\end{equation}
\section{Tsallis information dimension}
\label{Tsinformationd}
In this section, a Tsallis information dimension of the complex networks, ${d_T}$, is proposed as follows:

\begin{equation}
\label{TsiD}
{d_T}  = \frac{{\frac{{1 - \sum\limits_{i = 1}^N {pi(l)^q } }}{{q - 1}}}}{{\ln l}}(q \in \Re )
\end{equation}

Where $l$ is the box size in the box-covering algorithm. The numerator is the Tsallis entropy which is defined in Eq. (\ref{S_q1}). It can be easily seen that when $q=1$ the Tsallis information dimension is degenerated to the information dimension of complex networks in \cite{wei2014informationdimension}.

Similar to Shannon's information volume, we use the Tsallis entropy to define the information volume of complex networks as follows,

\begin{equation}
\label{Iv-2}
{I_v} = \frac{{1 - \sum\limits_{i = 1}^{{N_p}} {{p_i}^q} }}{{q - 1}}
\end{equation}

We discuss the relationship between the parameter ${q}$ and the information dimension of the complex networks.

 \textbf{CASE 1, when ${q \to -\infty}$}, the boxes with the minimum probability have the maximum effect on the information dimension of the complex networks.

 \textbf{CASE 2, when ${q \to 0}$}, the boxes with different probability have the same effect on the information dimension of the complex networks.

 \textbf{CASE 3, when ${q \to 1}$}, the Tsallis information dimension is degenerated to the information dimension in \cite{wei2014informationdimension}.

 \textbf{CASE 4, when ${q \to \infty}$}, the boxes with the maximum probability have the maximum effect on the information dimension of the complex networks. The information dimension of the complex networks is closed to 0.

It can be easily found that, with the increase of $q$, the information dimension of the complex networks is decreased.

%%%%%%%%%%%%%%%%%%%%%%%%%%%%%%%%%%%%%%%%%%%%%%%%%%%%%%%%%%%%%%%%%%%%%%%%%%%%%%%%%%%%%%%%%%%%%%

%Where ${d_T}$ is the Tsallis information dimension of complex networks, ${I_V}$ is defined in Eq. (\ref{Iv-2}) .

%%%%%%%%%%%%%%%%%%%%%%%%%%%%%%%%%%%%%%%%%%%%%%%%%%%%%%%%%%%%%%%%%%%%%%%%%%%%%%%%%%%%%%%%%%%%%

%So the value of $q$ will influence the Tsallis information dimension of the complex networks. The relationship is defined as follows:
%
%\begin{table}[!ht]
%\begin{center}
%\caption{The relationship between $q$ and  information dimension  }
%\label{tab-Tiq}
%\begin{tabular}{ll}
%\hline
%Scope of $q$ value&Tsallis information dimension \\
%\hline
%${q=0}$ & ${{d_T} = \mathop {\lim }\limits_{l \to 0} \frac{{{N_b} - 1}}{{\ln l}}}$ \\
%${0<q<1}$ & ${d_T}$ is controlled by those small value proability\\
%${q=1}$ & ${d_T}$ degenerate to classical information dimension \cite{wei2014informationdimension} \\
%${q>1}$ & ${d_T}$ is  controlled by those big value proability\\
%${q \rightarrow \infty}$ & The fractal dimension is close to 0 \\
%\hline
%\end{tabular}
%\end{center}
%\end{table}
%
%
%
%
%The Tsallis information dimension is a generalise information dimension of the complex networks. The information dimension of the complex networks is a dynamic quantity. It has revealed the fractal property of the complex networks from different aspects.
\section{Application}
\label{application}
In this section, we use the proposed method to calculate the information dimension of four real networks, namely, the US-airlines networks \cite{networkdata}, Email networks \cite{networkdata} and the Germany highway networks \cite{nettt}. The results are given in Table \ref{realnet1}.

\begin{table}[!ht]
  \centering
\caption{Tsallis information dimension of real networks}
\label{realnet1}
\begin{tabular}{lclc}
\hline
       & Germany highway \cite{nettt}  & Us-airline \cite{networkdata} &  Email \cite{networkdata}\\
       \hline
Nodes & 1168 & 500 & 1133 \\
edges & 2486 & 5962 & 10902 \\
 \hline
${d_{T(q = 0.1)}}$ & 61.88  & 62.89   & 175.20  \\
${d_{T(q = 0.5)}}$ & 10.646   & 15.780   & 23.778   \\
${d_{T(q = 1.0)}}$ & 1.9384   & 2.9682   & 3.5132  \\
${d_{T(q = 1.5)}}$ & 0.66732  & 1.0585  & 1.1131   \\
${d_{T(q = 2.0)}}$ & 0.35145   & 0.5758   & 0.5817   \\
${d_{T(q = 10)}}$ & 0,0268   & 0.0564   & 0.0562   \\
${d_{T(q = 100)}}$ & 0.0009   & 0.0037   & 0.0039  \\
${d_{T(q = 1000)}}$ & 0.0003   & 0.0002   & 0.00029 \\\hline
\end{tabular}
\end{table}

%\begin{table}[!h]
%\caption{Tsallis information dimension of real networks 1}
%\label{realnet1}
%\begin{tabular}{lllllll}
%\hline
%Networks        & Nodes & edages &    ${d_{T(q = 0.1)}}$      &    ${d_{T(q = 0.5)}}$      &  ${d_{T(q = 1.0)}}$      &${d_{T(q = 1.5)}}$\\ \hline
%%Protein interaction \cite{nettt}            & 2375  & 23386  & 398.33 & 23.206 & 2.9222 & 0.8953  \\
%Germany highway \cite{nettt} & 1168  & 2486   & 61.88 & 10.646 & 1.9384 & 0.66732 \\
%Us-airline  \cite{networkdata}    & 500   & 5962   & 62.89 & 15.780 & 2.9682 & 1.0585  \\
%Email      \cite{networkdata}     & 1133  & 10902  & 175.20 & 23.778 & 3.5132 & 1.1131  \\ \hline
%\end{tabular}
%\end{table}
%
%\begin{table}[!h]
%\caption{Tsallis information dimension of real networks 2}
%\label{ret2}
%\begin{tabular}{lllllll}
%\hline
%Networks        & Nodes & edages &    ${d_{T(q = 2.0)}}$    &   ${d_{T(q = 10)}}$     &   ${d_{T(q = 100)}}$ & ${d_{T(q = 1000)}}$   \\ \hline
%%Protein interaction \cite{nettt}           & 2375  & 23386  & 0.4645  & 0.0404 & 0.0023 & 0.00011     \\
%Germany highway \cite{nettt} & 1168  & 2486   & 0.35145 & 0.0269 & 0.0009 & 0.00030    \\
%Us-airline  \cite{networkdata}     & 500   & 5962   & 0.5758  & 0.0564 & 0.0037 & 0.00020    \\
%Email    \cite{networkdata}       & 1133  & 10902  & 0.5817  & 0.0562 & 0.0039 & 0.00029    \\ \hline
%
%\end{tabular}
%\end{table}

The comparison between the information dimension in \cite{wei2014informationdimension} and the proposed information dimension is shown in Table \ref{compare1}.

\begin{table}[!ht]
  \centering
\caption{The results with different methods }
\label{compare1}
\begin{tabular}{cllclll}
\hline
Networks        & Nodes & edages &    ${d_b}$    &   ${d_{T(q = 0.1)}}$     &    ${d_{T(q = 1)}}$ &  ${d_{T(q = 1000)}}$   \\ \hline
%Protein interaction \cite{nettt}          & 2375  & 23386  & 2.9222  & 398.33 &2.9222 & 0.00011     \\
Germany highway \cite{nettt} & 1168  & 2486   & 1.9384  & 61.8816 &1.9384 & 0.00030    \\
Us-airline \cite{networkdata}    & 500   & 5962   & 2.9682  & 62.8919 &2.9682 & 0.00020    \\
Email \cite{networkdata}     & 1133  & 10902  & 3.5132  & 175.21 &3.5132 & 0.00029    \\ \hline

\end{tabular}
\end{table}

In the Table \ref{realnet1} and Table \ref{compare1}, the ${d_b}$ represents the information dimension of the complex networks which is calculated by the method  in \cite{wei2014informationdimension}. The ${d_{T(q = x)}}$ represents the information dimension of complex networks which is calculated by the proposed method.

\begin{figure}[!ht]
  \centering
  \includegraphics[width=10cm]{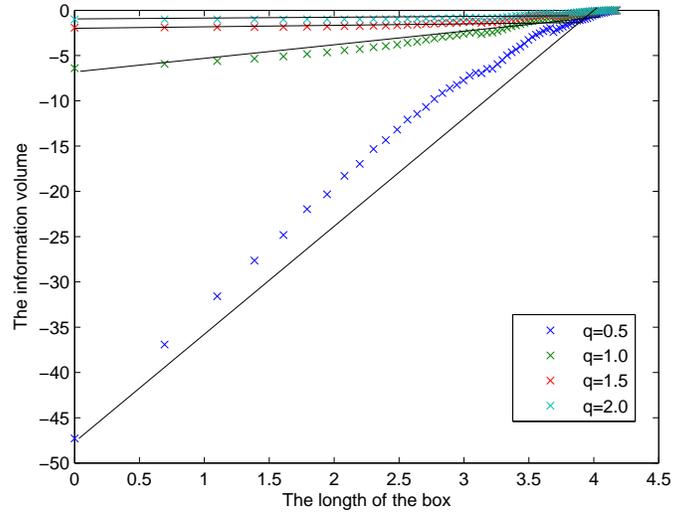}\\
  \caption{The Germany highway network}\label{highway}
\end{figure}

\begin{figure}[!ht]
  \centering
  \includegraphics[width=10cm]{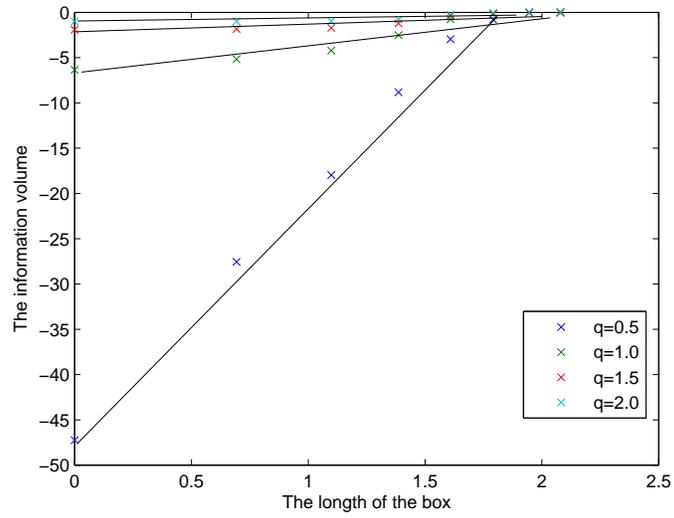}\\
  \caption{The Email network}\label{highway}
\end{figure}

\begin{figure}[!ht]
  \centering
  \includegraphics[scale=0.8]{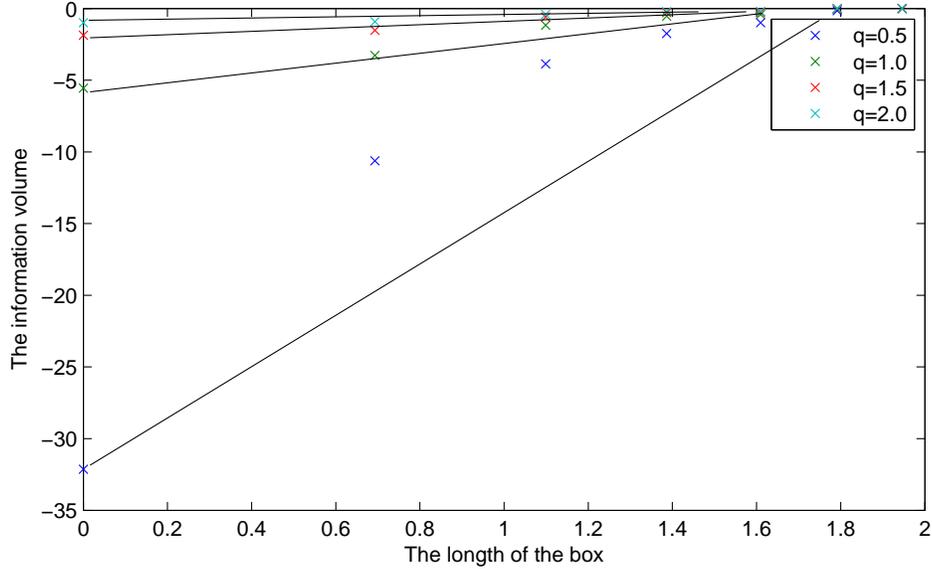}\\
  \caption{The US-airport network}\label{highway}
\end{figure}

% The dots which are shown in Fig. ги\ref{figemail}-\ref{figeUS-airline} with different colours have different coordinate.
%
% The y-coordinate of the dots represents the value of ${I_S}$ which is calculated by the proposed method based on different values of ${q}$ and different scale of the box size ${l}$.
% The x-coordinate of the dots is the value of ${\ln (l)}$.
%
The slope of the straight lines in the Figure (2-4) represents the information dimension of complex network. The results have shown that the information dimension is in inverse proportion to the value of ${q}$.
\section{Conclusion}
\label{conclusion}
The information dimension is widely used to illuminate the fractal and self-similarity properties of the complex networks. In this article, a general method to calculate the  information dimension of complex networks has been proposed based on the Tsallis entropy. It can be used to describe the influence of different parts in the complex networks to the fractal property. The proposed Tsallis information dimension is a generalization of the existing information dimension to the complex networks.
\section*{Acknowledgments}
The work is partially supported by National Natural Science Foundation of China (Grant No. 61174022), Specialized Research Fund for the Doctoral Program of Higher Education (Grant No. 20131102130002), R$\&$D Program of China (2012BAH07B01), National High Technology Research and Development Program of China (863 Program) (Grant No. 2013AA013801), the open funding project of State Key Laboratory of Virtual Reality Technology and Systems, Beihang University (Grant No.BUAA-VR-14KF-02).

%\appendix
%
%\section{1}

%% References


\bibliographystyle{elsarticle-num}
\bibliography{References}






% \end{thebibliography}

\end{document}